\documentclass[osajnl,twocolumn,showpacs,superscriptaddress,10pt]{revtex4-1} 
 \usepackage{amsmath,amssymb,graphicx}
\begin{document}

\title{Phase-and-amplitude recovery from a single phase contrast image using partially spatially coherent X-ray radiation}

\author{Mario A. Beltran}
\affiliation{Physics Department, Technical University of Denmark, DK-2800 Kgs, Lyngby, Denmark}
\affiliation{School of Science, RMIT University, Victoria 3001, Australia}
\affiliation{EMPA Swiss Federal Laboratories for Materials Science and Technology, 8600 D$\ddot{u}$bendorf, Switzerland}

\author{David M. Paganin}%
\affiliation{School of Physics and Astronomy, Monash University, Victoria, 3800, Australia }
 
\author{Daniele Pelliccia}
\affiliation{Instruments $\&$ Data Tools Pty Ltd, Victoria 3178, Australia}
\affiliation{School of Science, RMIT University, Victoria 3001, Australia}

\begin{abstract}
A simple method of phase-and-amplitude extraction is derived that corrects for image blurring induced by partially spatially coherent incident illumination using only a single intensity image as input. The method is based on Fresnel diffraction theory for the case of high Fresnel number, merged with the space--frequency description formalism used to quantify partially coherent fields and assumes the object under study is composed of a single material. $A$ $priori$ knowledge of the object's complex refractive index and information obtained by characterizing the spatial coherence of the source is required. The algorithm was applied to propagation-based phase contrast data measured with a laboratory-based micro-focus X-ray source. The blurring due to the finite spatial extent of the source is embedded within the algorithm as a simple correction term to the so-called Paganin algorithm and is also numerically stable in the presence of noise.   
\end{abstract}

\maketitle 

\section{Introduction} \label{Introduction}

Absorption-based X-ray radiography is a formidable non-invasive technique to study samples at micrometer and sub-micrometer length scales \cite{KakSlaney}. However, when the difference in absorption of the transmitted beam by different adjacent materials is small the specimen's features become difficult to visualize. Hence, absorption-based X-ray radiography has limited applications, when imaging weakly absorbing biological specimens without contrast agents \cite{Zernike}. 

One way to overcome this limitation is phase-contrast imaging (PCI). In PCI the phase shifts undergone by the transmitted X-rays as they traverse the sample are converted into transverse intensity variations. Various PCI techniques exist. Examples are X-ray interferometry \cite{BonseHart}, analyzer-based phase contrast \cite{Foster}, X-ray grating methods \cite{PfeifferNature}, and propagation-based phase-contrast imaging (PBI) \citep{{Gabor},WilkinsNature}. Each have their relative advantages and disadvantages.

\begin{figure}[h]
\centering
\includegraphics[scale=0.4]{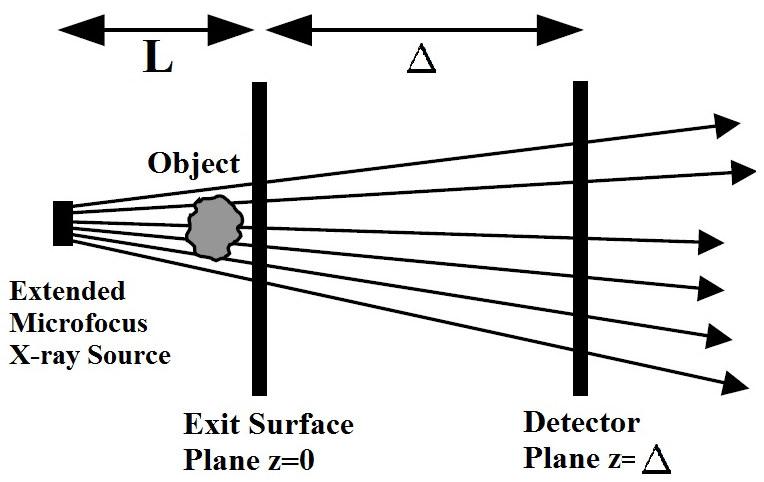}
\caption{Schematic diagram of a propagation-based X-ray phase-contrast imaging setup with an extended source.} 
\label{Fig:MicroFocusPBISetUp}
\end{figure}

Here, we limit ourselves to PBI as it solely relies on free-space propagation to render contrast thus making it simple to implement \cite{WilkinsNature}. PBI images can be measured by placing a position-sensitive detector at some non-zero distance downstream from the exit surface of the object (see Fig.~\ref{Fig:MicroFocusPBISetUp}). While sources with low temporal coherence can be utilized, sufficient spatial coherence (e.g. through a sufficiently small source size) is typically needed to yield significant contrast in such PBI images \cite{WilkinsNature}.

To infer quantitative information, phase-retrieval methods can be applied to the raw phase contrast intensity measurements \cite{GerchbergSaxton}. Most phase-retrieval methodologies require two or more intensity images and also often rely on the assumption that the incoming radiation is fully coherent \citep{GerchbergSaxton,TeagueTIEPaper}. In the past decades several phase retrieval algorithms have been developed that enable the extraction of phase-and-amplitude information from a single PBI intensity measurement \citep{PaganinAlg,TurnerAlg,BeltranAlg}. These methods were derived under the assumption that the incident radiation is fully coherent. In X-ray PBI, the degree of spatial coherence of the incident beam has significant consequences on the intensity images. Basically, the visibility of phase contrast fringes is directly affected by the spatial coherence, that is, lower-spatial-coherence beams proportionally degrade fringe visibility \cite{Myers2009}. With PBI setups using lower coherence laboratory-based X-ray sources becoming more routinely used nowadays it is becoming progressively more important that coherence effects be accounted for in phase-and-amplitude extraction methods. 

In this paper we derive and experimentally verify a simple yet robust algorithm that recovers the projected thickness of single-material specimens that considers the loss of fringe visibility due to the partial spatial coherence from the emitting source, whilst still only using one PBI intensity image as input. The rectification due to partial coherence comes as a simple correction term within the algorithm itself which can be obtained by prior characterization of the source. Similar to methods presented previously, it requires $a$ $priori$ knowledge of the absorptive and refractive properties of the imaged object. Our derivation ultimately leads to a mildly adjusted form of the so-called Paganin method \cite{PaganinAlg}. This modification consists of replacing the object-to-detector propagation distance with an effective propagation distance so as to partially deconvolve the smearing effect of finite source size, in a numerically stable manner.  

A simple means for both motivating and conceptualising our key finding is as follows. The Paganin method, in essence, applies a low-pass Fourier-space filter to a single propagation-based phase contrast image in order to yield a projected thickness map of a single-material object. Therefore, if the finite source size has already blurred the image somewhat, the degree of blurring required by the Paganin filter should be reduced. As we shall see via the calculation presented below, the effects of the partial spatial coherence may be accounted for by reducing the so-called delta-to-beta ratio which parametrizes the Paganin filter, this reduction being by an amount proportional to the area of the source. This effects a simple and stable partial source-size deconvolution in the recovered thickness map of the single-material object.

\section{Theory} \label{Theory}

We begin our derivation by considering Fig.~\ref{Fig:MicroFocusPBISetUp}. For the moment, assume monochromatic X-rays that are traveling parallel to the optic $z$-axis. The complex field $\Psi (\textbf{r},\Delta )$ formed at the detector plane ($z=\Delta$) will be given by the Fresnel diffraction integral \cite{Guigay}. However, since the distance $\Delta > 0$ considered here is assumed small enough such that only a single phase-contrast fringe is visible about regions of localized boundaries then the integral will reduce to the following form, corresponding to a Fresnel number that is much greater than unity \cite{PaganinBook}:
 
\begin{eqnarray} 
\Psi  (\textbf{r},\Delta)= \left ( 1+\frac{i\Delta }{2k}\nabla^{2}_{\textbf{r}} \right ) \Psi^{0} (\textbf{r}). 
\label{eqn:ParCohTIE1}
\end{eqnarray}

\noindent Here, $\Psi^{0} (\textbf{r})$ is the field at the exit surface plane $z=0$, $\textbf{r}=(x,y)$ are coordinates transverse to the optic axis $z$, and $\nabla^{2}_{\textbf{r}} = \partial^{2} / \partial x^{2}+\partial^{2} / \partial y^{2}$ is the transverse Laplacian operator. Note that the above expression may also be obtained by taking the parabolic equation of paraxial scalar wave optics, and applying a two-point finite-difference approximation to the $z$-derivative of the field.

To incorporate the effects of partial coherence we utilize the space--frequency description due to Wolf~\cite{WolfSpaceFreqPaper}, whereby partially coherent fields are described in terms of the cross-spectral density obtained at each temporal frequency via statistically averaging over an ensemble of strictly monochromatic fields, all of which have that same temporal frequency: 
 
\begin{eqnarray} 
W(\textbf{r}_{1 },\textbf{r}_{2},\Delta)=\left \langle \Psi_{\theta}^{*} (\textbf{r}_{1},\Delta)\Psi_{\theta} (\textbf{r}_{2 },\Delta) \right \rangle _{\theta} . 
\label{eqn:ParCohTIE2}
\end{eqnarray}

\noindent The angular brackets $\left \langle  \right \rangle _{\theta} $ denotes the ensemble average of all possible fields with direction given by the vector $\textbf{k}_{\theta}$, since we work with a plane-wave ensemble by assumption. Note, $\theta$ is the angle $\textbf{k}_{\theta}$ makes with respect to the $z$-axis. One can readily obtain expressions for $\Psi^{*}_{\theta} (\textbf{r}_{1 },\Delta)$ and $\Psi_{\theta} (\textbf{r}_{2 },\Delta)$. Substituting these into Eq.~\ref{eqn:ParCohTIE2} followed by setting $\textbf{r}_{1}=\textbf{r}_{2}=\textbf{r}$ yields the intensity at the detector plane $z=\Delta$:

\begin{eqnarray} 
\textup{I}(\textbf{r},\Delta)= \left \langle \left | \Psi^{0}_{\theta}  (\textbf{r}) \right |^{2} \right \rangle _{\theta}  +\frac{\Delta }{k} \textup{Re}\left \{ i \left \langle \Psi^{0*}_{\theta} (\textbf{r}) \nabla^{2}_{\textbf{r}} \Psi^{0}_{\theta}  (\textbf{r}) \right \rangle _{\theta}  \right \}. \nonumber\\
\label{eqn:ParCohTIE4}
\end{eqnarray}

From this point we only consider objects comprised of a single material, that is, those whose complex refractive index $n=1-\delta+i\beta$ is constant throughout the volume \cite{PaganinAlg}. More precisely, we assume that the ratio of $\delta / \beta$ is everywhere the same within the volume occupied by the sample, thereby permitting it to be of variable density. Under this approximation the exit surface wavefield $\Psi^{0}_{\theta}(\textbf{r})$ for an arbitrary angular orientation $\theta$ can be described with the projection approximation \cite{PaganinBook}:

\begin{eqnarray} 
\Psi^{0}_{\theta} (\textbf{r})=\sqrt{\textup{I}^{\textup{IN}}} e^{- \left ( \frac{\mu}{2}+ik\delta \right ) \textup{T}_{\theta}(\textbf{r}) }, \nonumber\\
\label{eqn:ParCohTIE5}
\end{eqnarray}
 
\noindent where $\textup{T}_{\theta}(\textbf{r})$ is the sample's projected thickness for a particular $\theta$. $\textup{I}^{\textup{IN}}$ is the intensity of the incident of the beam before it impinges on the sample. $\mu=2k\beta$ is the linear attenuation coefficient at wavelength $\lambda=2\pi/k $. Upon substitution into Eq.~\ref{eqn:ParCohTIE4} the intensity at $z=\Delta$ becomes 

\begin{eqnarray} 
\textup{I}(\textbf{r},\Delta) = \textup{I}^{\textup{IN}} \left (  1-\frac{\delta \Delta }{\mu }\nabla^{2}_{\textbf{r}} \right ) \left \langle e^{-\mu \textup{T}_{\theta}(\textbf{r})} \right \rangle _{\theta} . \nonumber\\
\label{eqn:ParCohTIE7}
\end{eqnarray}
 
Before proceeding we draw attention to the term on the right-hand-side that describes an ensemble average of transmitted intensities (Beer's law) at various $\theta$ values. From a purely parallel-ray projection perspective it is known that such averaging of transmitted intensities will be manifested as penumbral blur in the registered image \cite{Myers2009}. Suppose this image blur can be approximated as a convolution of the incident exit surface intensity at $\theta=0$ (i.e. projection along the optic axis $z$) with a Gaussian:


\begin{eqnarray} 
 \left \langle e^{-\mu \textup{T}_{\theta}(\textbf{r})} \right \rangle _{\theta}&=&\exp\left ( -\frac{\textbf{r}^{2}}{2\sigma ^{2} }  \right ) \otimes e^{-\mu \textup{T}(\textbf{r})}. \nonumber\\
\label{eqn:ParCohTIEConvolution}
\end{eqnarray}

\noindent  Here, $\textup{T}(\textbf{r})$ is the projected thickness along the $z$-axis (i.e. $\theta=0$), $\otimes$ denotes convolution, and $\sigma=\sqrt{ \sigma_{x}^{2}+\sigma_{y}^{2}}$ is the standard deviation of the Gaussian function. Note, if the Gaussian function is symmetric about the $x-y$ plane, then $\sigma_{x}^{2}=\sigma_{y}^{2}$. We assume this to be the case, but note that our formulae are readily modified if this approximation is not applicable. Invoking the Fourier convolution theorem followed by Taylor approximating the Fourier space Gaussian leads to the expression below:    

\begin{eqnarray} 
 \left \langle e^{-\mu \textup{T}_{\theta}(\textbf{r})} \right \rangle _{\theta} &&=\mathcal{F}^{-1} \left [  \exp \left(-2 \sigma |\textbf{k}_{\textbf{r}}|^{2} \right)   \mathcal{F} \left \{  e^{-\mu \textup{T}(\textbf{r})} \right \}\right ] \nonumber\\
&&\approx  \mathcal{F}^{-1}\left [(1-2\sigma ^{2} |\textbf{k}_{\textbf{r}}|^{2})   \mathcal{F} \left \{  e^{-\mu \textup{T}(\textbf{r})} \right \}\right ]. \nonumber\\
\label{eqn:ParCohTIEConvolutionSTEP}
\end{eqnarray}

\noindent Here, $\textbf{k}_{\textbf{r}}= (k_{x},k_{y})$ are Fourier space coordinates dual to $\textbf{r}=(x,y)$, while $\mathcal{F} $ and $\mathcal{F}^{-1}$ denote forward and inverse Fourier transforms. Substituting Eq.~\ref{eqn:ParCohTIEConvolution} into Eq.~\ref{eqn:ParCohTIE7}, then making use of the Fourier derivative theorem and neglecting any terms higher than $|\textbf{k}_{\textbf{r}}|^{2}$ in order, one can then solve for object's projected thickness $\textup{T}(\textbf{r})$ via:

\begin{eqnarray} 
\textup{T}(\textbf{r})&&= \nonumber\\
&&-\frac{1}{\mu} \textup{ln}  \left [ \mathcal{F}^{-1}  \left ( \frac{1}{ 1 +  \left ( \frac{\delta \Delta}{\mu} - 2\sigma ^{2} \right ) |\textbf{k}_{\textbf{r}}|^{2} } \mathcal{F} \left \{ \frac{\textup{I}(\textbf{r},\Delta)}{\textup{I}^{\textup{IN}}} \right \} \right ) \right ]. \nonumber\\ 
\label{eqn:ProjectedThicknessRecovery}
\end{eqnarray}

This is identical to the commonly-used algorithm of Paganin~$et$~$al$.~\cite{PaganinAlg}, the only difference being that blurring effect of finite source size is compensated by replacing the propagation distance $\Delta$ with the effective distance (cf. the deblur by defocus concept introduced by Eq. 27 of Gureyev~$et$~$al$.~\cite{Gureyev2006opt}):

\begin{eqnarray} 
\Delta_{\textrm{eff}}=\Delta -2\sigma^{2} \mu / \delta .
\label{eqn:DistanceEffective}
\end{eqnarray}

An alternative way of expressing this, obtained by recalling that $\mu=2 k \beta$ is to consider the ratio $\delta / \beta$ of the object to be replaced by the effective ratio:

\begin{eqnarray} 
\left (  \frac{\delta }{\beta}  \right )_{\textup{eff}}=\frac{\delta }{\beta} -\frac{8 \pi \sigma^{2}}{\lambda \Delta}.
\label{eqn:RatioEffective}
\end{eqnarray}

Interestingly, in Eqs.~\eqref{eqn:ProjectedThicknessRecovery} through to \eqref{eqn:RatioEffective} the correction term is proportional to the area of the source. In fact, more generally, $\sigma$ in Eq.\eqref{eqn:RatioEffective} may be taken as the total blurring width due to the combined effect of finite source size, detector-induced smearing, etc.  

So far we have ignored the beam's cone-like geometry. For the imaging setup used here the size of the source is several times smaller than the size of the object, allowing us to use simple point-projection geometry to account for image magnification $\textup{M}$. This factor will be given by $\textup{M}=(\textup{L}+\Delta)/\textup{L}$ and can be trivially inserted into Eq.~\ref{eqn:ProjectedThicknessRecovery} as was done in Paganin~$et$~$al$.~\cite{PaganinAlg}. 

\section{Experimental Results} \label{Results}

\begin{figure}[h]
\centering
\includegraphics[scale=0.54]{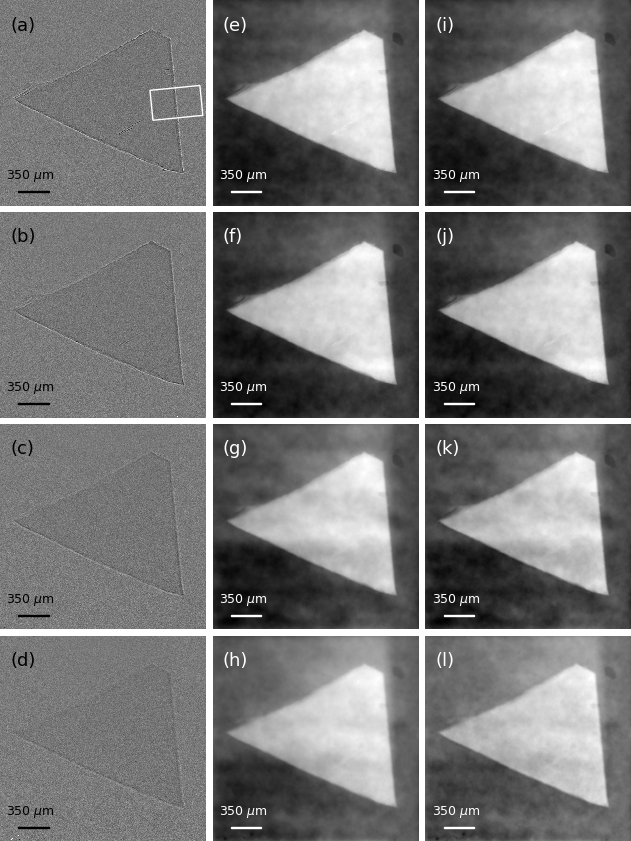}
\caption{(a-d) PBI images of the Kapton taken at a fixed source-to-detector distance with different source sizes to vary the spatial coherence. (e-h) display images of the retrieved projected thickness from (a-d), respectively, using Eq.~\ref{eqn:ProjectedThicknessRecovery} ignoring the blurring correction term $2\sigma ^{2}$. (i-l) display images of the retrieved projected thickness as in (e-h), however, on this occasion the term $2 \sigma ^{2}$ for the corresponding different sources sizes is included.} 
\label{Fig:PhaseRetResults}
\end{figure}

To validate Eq.~\ref{eqn:ProjectedThicknessRecovery} X-ray propagation-based phase-contrast experiments were performed using a laboratory based micro-focus X-ray source. Here, X-rays are generated using a standard voltage vacuum tube that focuses a beam of electrons onto a tungsten target. For all the images acquired, the voltage of the tube was kept constant at 65 kV  producing a polychromatic X-ray beam with a maximum photon energy of 65 keV. A CCD camera was positioned at a distance $\textup{L}+\Delta= 126$ cm from the source to measure the images. Flat and dark field recordings were acquired for every set to correct for the beam's non-uniformity and electronic noise. The object was placed at a distance $\textup{L}=15$ cm from the source. Based on the point-projection geometry discussed earlier, this setup gave magnification $\textup{M}=5.7$ and an effective pixel size of 2.3 $\mu$m. As a test sample we used 50 $\mu$m thick Kapton film ($\textup{C}_{22} \textup{H}_{10} \textup{N}_{2} \textup{O}_{5}$) cut into a roughly triangular shape.  

PBI images with different states of X-ray beam coherence were taken. To vary the spatial coherence the current of the focused electron beam was altered to change the source size. To keep photon fluence fixed for every PBI image taken at a particular current setting we altered the exposure time accordingly, so that for each data set the total number of counts was approximately equal. The value of $\sigma^2$ for each current setting has been experimentally determined by measuring the width of the slope of the image of the edge of a pinhole (measurements not shown). The projection of a straight edge will be blurred to an amount that is proportional to the source size. By knowing the geometry of the image, we estimated the source size to be 3.9 $\mu$m, ​5.​8 $\mu$m​, ​8.​6 $\mu$m​, and ​10.​6 ​​$\mu$m​ for the ​​​​50, 100, 150 and 200 ​​$\mu$​A current setting respectively.  

\begin{figure}[h]
\centering
\includegraphics[scale=0.5]{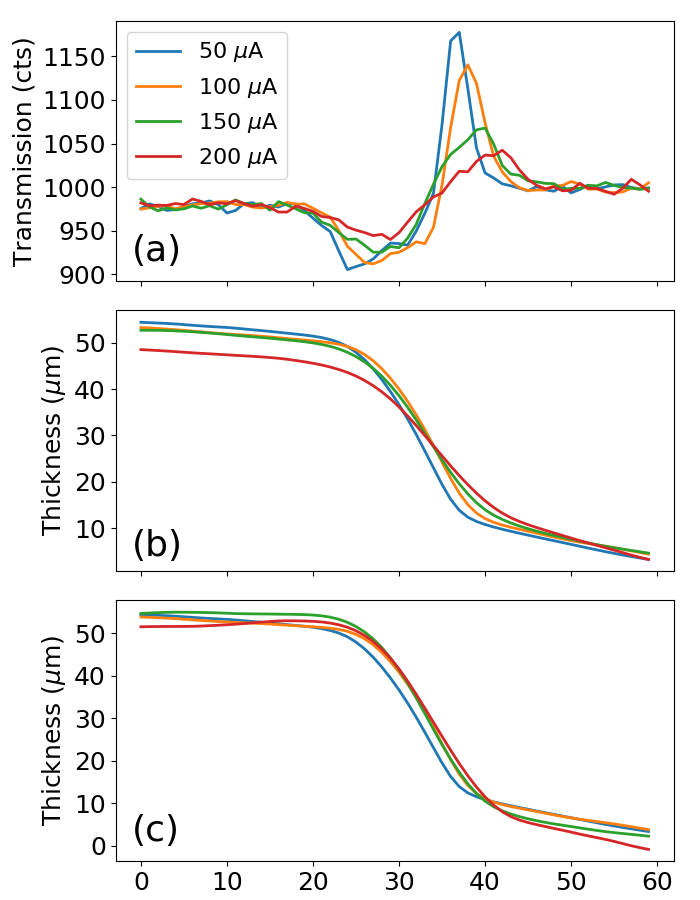}
\caption{(a) Localized line profile plots of the same regions in Fig.~\ref{Fig:PhaseRetResults} (a-d). (b) Localized line profile plots of the same regions in Fig.~\ref{Fig:PhaseRetResults} (e-h). (c) Localized line profile plots of the same regions in Fig.~\ref{Fig:PhaseRetResults} (i-l). Note, the marked box in Fig.~\ref{Fig:PhaseRetResults} (a) also corresponds to the region where all the other profiles were taken.} 
\label{Fig:Profiles}
\end{figure}

Figures~\ref{Fig:PhaseRetResults} (a-d) show X-ray PBI images of the Kapton object at different electron beam currents. The current setting for the images in (a-d) were as respectively stated previously. In all four PBI images fringes resulting from near field Fresnel diffraction are clearly observed at the air/object boundaries. It is evident from these images that at higher current settings the visibility of the object features reduces as result of the decrease in spatial coherence of the X-ray beam. This is further highlighted in Fig.~\ref{Fig:Profiles} (a) where overlaid localized line profiles of the raw phase-contrast image of the sample's edge marked by the boxed region for the different currents are presented. 

Figures~\ref{Fig:PhaseRetResults} (e-h) show the recovered projected thickness of the object via the $\sigma=0$ limit of Eq.~\ref{eqn:ProjectedThicknessRecovery} from their corresponding respective PBI images in (a-d). Since the X-ray beam is polychromatic, effective values of the complex refractive $\delta$ and $\mu$ specific for Kapton were used \cite{Arhatari}. The value for $\mu=386.7$ $m^{-1}$ was estimated using a specific region from the raw PBI images via Beer's law. The region was chosen such that any signal was solely due to absorption (i.e. away from fringes). The value $\delta=3.38 \times 10^{-7}$ was obtained from data available at http://www.nist.gov/index.html. As previously mentioned, for the set of retrieved images (e-h) the correction factor $2\sigma ^{2}$ due to source size blurring was not included in Eq.~\eqref{eqn:ProjectedThicknessRecovery}. Omission of this term results in an over-smoothing in the reconstructed projected thickness images, in particular if those reconstructions come from PBI images taken with lower spatial coherence (i.e. larger source size). This over-smoothing originates from an excess Fourier filtration of high spatial frequencies resulting from setting $\sigma$ to zero in Eq.~\ref{eqn:ProjectedThicknessRecovery} \cite{BeltranAlg}.

In Fig.~\ref{Fig:PhaseRetResults} (i-l) we show the recovered projected thickness of the object using Eq.~\ref{eqn:ProjectedThicknessRecovery} from their corresponding PBI images in (a-d). However, now the correction term $2\sigma ^{2}$ corresponding to each current setting is included for each calculation (see Eq.~\eqref{eqn:ProjectedThicknessRecovery}). By the inclusion of this factor it is expected that all retrieved images (i-l) display a somewhat similar quality on account of properly adjusting the denominator of the Fourier filter by $\delta \Delta/ \mu +2\sigma ^{2}$. 

To quantitatively compare the results in Fig.~\ref{Fig:PhaseRetResults}, line profiles are presented in Fig.~\ref{Fig:Profiles}. In Fig.~\ref{Fig:Profiles} (b) overlaid profiles of the same region in Fig.~\ref{Fig:PhaseRetResults} (e-h) are displayed. The blue, green, red and light-blue profiles in Fig.~\ref{Fig:Profiles} (e) correspond to retrieved images in (e-h), respectively. Here, we see that for the most part the profiles largely differ from each other especially near the top- and bottom-edge regions of the curve. On the other hand when overlaid profiles of the same region of the retrieved images that contain the proper correction factor, namely Fig.~\ref{Fig:PhaseRetResults}  (i-l) are displayed (see Fig.~\ref{Fig:Profiles} (c)) we see that all the curves get sharper at the edges and also begin to converge. Again this is more evident in the top-edge regions of the curves.

\section{Discussion and Conclusion} \label{Conclusion}

It is evident from the images in Fig.~\ref{Fig:PhaseRetResults}, that the algorithm performs well under the presence of noise in the input image. Having arrived at a slightly altered form of the algorithm in Paganin $et$ $al$.~\cite{PaganinAlg}, it is no surprise that the method presented has also inherited the same numerical stability \cite{GureyevNes2017}. Incidentally, the blurring correction term in the denominator also reveals limitations of the method. For example, if the source size is large enough such that $2 \sigma^{2}>\delta \Delta / \mu $ then certain $\textbf{k}_{\textbf{r}}$ values will yield zeroes in the denominator and therefore give rise to instabilities. To ensure this does not occur the size of the source must not be too large so that the degree of spatial coherence is sufficiently high to produce a sizable single phase-contrast fringe in the data. Indeed, requiring that the actual and effective propagation distances be close to one another leads to the rule of thumb that the source area $A=\pi \sigma^{2}$ obey $A << \pi \delta \Delta/ (2 \mu)$. Another feature worth mentioning is that one may adjust the propagation distance by an amount $\Delta \rightarrow \Delta +2 \sigma^{2} \mu / \delta$ and only input the values of $\delta$ and $\mu$ and obtain the same result. From a signal processing view point the excess high-spatial frequency suppression caused by excluding $2\sigma^{2}$ may be compensated by increasing the propagation distance to enlarge the size of the fringe as a high-spatial frequency signal boosting mechanism.

In this paper we have derived a simple and practical algorithm for retrieving phase-and-amplitude information from a single propagation-based phase contrast image. The algorithm is not restricted to monochromatic and spatially coherent radiation and was successfully applied to data taken from a laboratory-based X-ray phase contrast point-projection microscope at different states of spatial coherence. Although this study considered an imaging system with only propagation-induced contrast, this idea of partial spatial coherence rectification can also be transferred to other phase contrast optical setups such as crystal-analyzer and grating-based (periodic and random) systems \citep{Pavlov2004,DiemozLondon,Morgan2011,Morgan2012}. The correction term comes as a simple adaptation of the method of Paganin $et$ $al$.~\cite{PaganinAlg} and can be easily be incorporated in freely available image processing software. The ImageJ plugin written by Weitkamp $et$ $al$.~\cite{WeitkampANKA} is one alternative. Finally, we anticipate that this technique will find use in microscopic systems that utilize other types of radiations as a probe (e.g. visible light, electrons and neutrons).

We close by recalling that many workers use the Paganin filter in a phenomenological manner, tuning the ratio $\delta / \mu$ of until the most acceptable reconstruction is obtained. Here, ``most acceptable'' may be equated to ``sharpest reconstruction with acceptably small artefacts''. The results of the present paper imply that such an approach, when applied to propagation-based phase-contrast data that satisfy the validity conditions specified in our calculation, achieves an implicit approximate source-size and detector-smearing deconvolution.

\section{Acknowledgements}

M. A. Beltran dedicates this paper to the memory of Maria Rosenda Herrera Donoso. D. M. Paganin acknowledges useful discussions with T. E. Gureyev.

\end{document}